\documentclass{mpe_report}
\usepackage{psfig}
\usepackage{epsfig}
\usepackage{natbib}

\begin{document}
\title{New insights into the X-ray properties of the nearby barred spiral galaxy NGC~1672}

\author{L. P. Jenkins\inst{1}, W. N. Brandt\inst{2}, E. J. M. Colbert\inst{3}, A. J. Levan\inst{4}, T. P. Roberts\inst{5}, M. J. Ward\inst{5}, A. Zezas\inst{6}}
\institute{Laboratory for X-ray Astrophysics, NASA Goddard Space Flight Center, Code 662, Greenbelt, MD 20771, USA
\and  Department of Astronomy \& Astrophysics, The Pennsylvania State University, University Park, PA 16802, USA
\and Johns Hopkins University, 3400 N. Charles St, Baltimore, MD 21218, USA
\and Department of Physics, University of Warwick, Coventry, CV4 7AL, UK
\and Department of Physics, Durham University, South Road, Durham, DH1 3LE,  UK
\and Harvard-Smithsonian Center for Astrophysics, 60 Garden Street, Cambridge, MA 02138, USA
}

\authorrunning{L.P. Jenkins et al.}
\titlerunning{New insights into the X-ray properties of NGC1672}

\maketitle

\begin{abstract}

We present some preliminary results from new {\it Chandra} and {\it XMM-Newton} X-ray observations of the nearby barred spiral galaxy NGC~1672. It shows dramatic nuclear and extra-nuclear star formation activity, including starburst regions located near each end of its strong bar, both of which host ultraluminous X-ray sources (ULXs). With the new high-spatial-resolution {\it Chandra} imaging, we show for the first time that NGC~1672 possesses a faint ($L_X\sim10^{39}$\,erg s$^{-1}$), hard central X-ray source surrounded by an X-ray bright circumnuclear starburst ring that dominates the X-ray emission in the region. The central source may represent low-level AGN activity, or alternatively the emission from X-ray binaries associated with star-formation in the nucleus. 

\end{abstract}

\section{Introduction}

The presence of a bar in a galaxy plays an important role in its evolution.  Approximately two-thirds of all spiral galaxies are barred \citep{sheth07}, so any complete picture of galaxies must include the impact of bars on their properties.   For example, there is believed to be a causal connection between the existence of a bar and a galaxy's circumnuclear properies.  Bars can drive gas from the disc to the inner regions as gas clouds lose angular momentum as they encounter the bar and sink towards the galaxy's centre.  Indeed, a substantial fraction of galaxies with bars do show enhanced star-formation activity in their central regions, some of which are also known to have circumnuclear rings ($\sim$20\%, \citealt{knapen05}).  

A question that we are addressing with this study is: can the material driven to the centre of a galaxy by the bar also fuel an active galactic nucleus (AGN)?  At present, optical studies show no significant correlation between Seyfert activity and bars (e.g. \citealt{ho97}; \citealt{knapen05}).  However, an X-ray study by \cite{maiolino99} shows that there is a strong correlation between absorbing column density toward Type-2 Seyfert nuclei and the presence of a strong bar: {\it more than 80\% of Compton thick Seyfert 2s are barred}.  This suggests that even if low-luminosity AGN activity is present, it may be obscured from view at optical wavelengths and/or diluted by strong star-formation activity, which could lead to a misclassification as a pure star-forming nucleus.

The topic of the influence of bars on galaxy evolution is not well studied at X-ray energies, but we are now able to utilize the excellent complementary imaging and spectral capabilities of {\it Chandra} and {\it XMM-Newton} to make detailed investigations of the X-ray characteristics of these systems.  X-ray studies offer a powerful means of probing star-formation activity in galaxies, i.e. from X-ray binaries (XRBs), supernova remnants (SNRs) and hot diffuse gas.  X-ray observations are also ideal for searching for obscured AGN activity in the form of a hard central point source that may not be visible at other wavelengths.  

To address the nature of the nuclear and off-nuclear X-ray sources in barred galaxies, we have obtained new X-ray observations of NGC~1672 with both {\it Chandra} (40\,ks) and {\it XMM-Newton} (50\,ks).   Here we show some preliminary results from these observations; a full analysis will be presented in Jenkins et al. (2008) {\it in prep}.

\begin{figure*}
\begin{center}
\scalebox{0.8}{\includegraphics{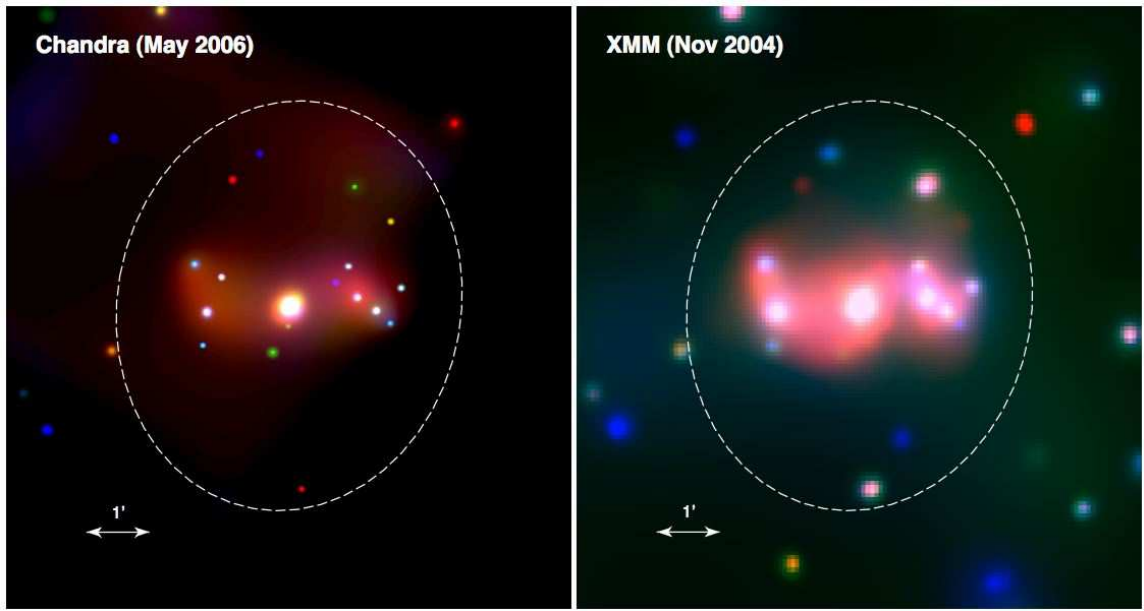}}
\caption{{\it Chandra} ACIS-S (left) and {\it XMM-Newton} EPIC (right) 3-color images of NGC~1672. Red=0.3--1\,keV, Green=1--2\,keV \& Blue=2--10 keV emission. North is up, east to the left.}
\label{fig:rgb}
\end{center}
\end{figure*}

\section{NGC~1672}

For this pilot study, we have selected the nearby (16.3\,Mpc) late-type barred spiral galaxy NGC~1672 (SB(r)bc).  It is known to have a strong bar (2.4 arcmin in length), and vigorous star formation is seen at its ends.  It has a high infrared (IR) luminosity (log $L_{FIR}/L_{\odot} = 10.4$), and many H{\rm II} regions in its four spiral arms (\citealt{brandt96}, and references therein).  

Optical studies have given conflicting evidence as to the nature of the nuclear activity in NGC~1672.  It was noted by \cite{sersic65} as having a complex or `amorphous' nuclear morphology, and was subsequently shown by \cite{storchi96} to possess a circumnuclear ring of star formation, measuring {$\sim11\times9$ arcseconds, located between two inner Lindblad resonances (ILRs). Therefore, some studies of its nuclear properties that use large apertures are strongly affected by the star-formation in the ring.  For example, \cite{osmer74} showed that the optical emission spectrum in the central 20 arcsec was similar to those of normal H{\rm II} nuclei, but with an H$\alpha$/H$\beta$ ratio indicative of a large amount of dust reddening.  \cite{storchi95} also measured optical emission-line ratios in a relatively large $10\times20$ arcsec$^2$ aperture, and classed the nucleus as `composite', with ratios between those of typical starburst and Seyfert values. More recent optical line ratio diagnostics by \cite{kewley00} of the emission within a $\sim13$ arcsec slit are indicative of an H{\rm II}-type nucleus, with no H$\alpha$ line broadening detected.  All of these studies will have been strongly contaminated by the emission from the circumnuclear starburst ring; however they do demonstrate that this component is dominant, and that any potential AGN activity would be relatively weak. 

However, more spatially-resolved studies have also given ambiguous results.  \cite{veron81} detected possible broadening of [O{\rm III}] ($\sim$300\,km s$^{-1}$) compared to H$\beta$ lines ($\sim$150\,km s$^{-1}$) in the central $2\times4$ arcsec$^2$, which they suggested was evidence of a composite H{\rm II}/Seyfert~2 nucleus. \cite{garcia90} performed a high-spatial-resolution spectral analysis of the nucleus, and found a strong increase in the [O{\rm III}]/H$\beta$ ratio in the central 1 arcsec compared to its immediate surroundings.  However, both lines had the same FWHM of $\sim$300\,km s$^{-1}$ and the authors classified the nucleus as a LINER.  \cite{storchi96} also measured spatially-resolved (2 arcsec) emission-line ratios, which were better modeled by a LINER-type stellar photoionization model with $T_{eff}\geq 45,000$\,K, rather than photoionization by a strong AGN continuum.

X-ray studies offer an excellent diagnostic for nuclear activity that may be obscured, and NGC~1672 has previously been observed with {\it ROSAT} and {\it ASCA}.  Three bright X-ray sources were detected in the soft 0.2--2.0\,keV band with the {\it ROSAT} HRI that were clearly associated with the galaxy (\citealt{brandt96}; \citealt{denaray00}). The brightest source was located at the nucleus, and the other two were cospatial with the ends of the bar.  The nucleus had a soft X-ray spectrum consistent with thermal emission with a temperature of 0.68\,keV, with a soft X-ray luminosity (0.2--2.0\,keV) of $7\times10^{39}$\,erg s$^{-1}$.  This is very weak compared to a typical Type-2 AGN, but within the range observed for starburst/H{\rm II} region galaxies.  However, these studies were again unable to resolve the nuclear emission spatially due to the 5 arcsec FWHM of the HRI, and so could not determine whether the bulk of the soft emission came from a starburst or hidden AGN nucleus.

\section{New X-ray Data}

\begin{figure*}
\begin{center}
\scalebox{0.8}{\includegraphics{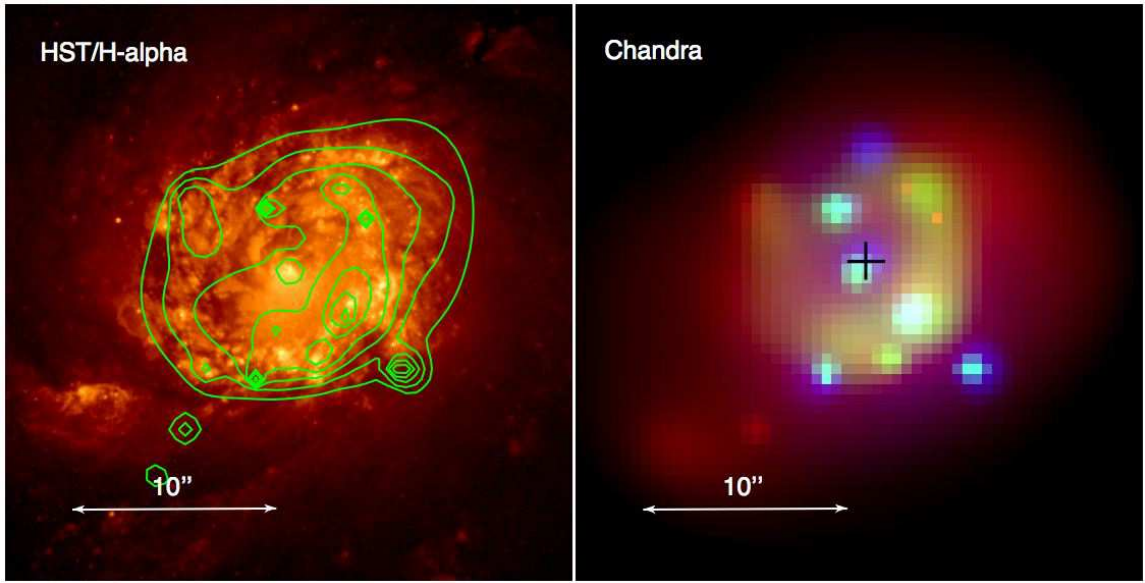}}
\caption{Left: {\it HST}/ACS H$\alpha$ image of the central region of NGC~1672, overlaid with green contours from the {\it Chandra} data.  A nuclear spiral is evident in this image.  Right: a {\it Chandra} 3-color X-ray image on the same scale (colors as in Fig.~1).  The circumnuclear ring is evident, surrounding a weak X-ray point source (marked with a cross).}
\label{fig:ctr}
\end{center}
\end{figure*}

Figure~\ref{fig:rgb} shows 3-color images (0.3-10\,keV) of the {\it Chandra} data ({\it left}) and {\it XMM} data ({\it right}).  The data show numerous point sources in the spiral arms and at both ends of the bar, which, if associated with NGC~1672, are likely to be XRBs/SNRs.  Note that the {\it XMM} data are also able to resolve the point sources very well.  A large number of these sources are brighter than $10^{39}$\,erg s$^{-1}$ at the distance of NGC~1672 (i.e. ultraluminous X-ray sources: ULXs), and are mostly clustered around the bar ends/inner spiral arms, including the two brightest extra-nuclear sources detected with {\it ROSAT}.  Soft diffuse emission is also seen tracing the bar structure (visible as soft red emission in both images).  A full analysis of the point-source populations and diffuse emission will be presented in Jenkins et al. (2008) {\it in prep}.

\begin{figure*}
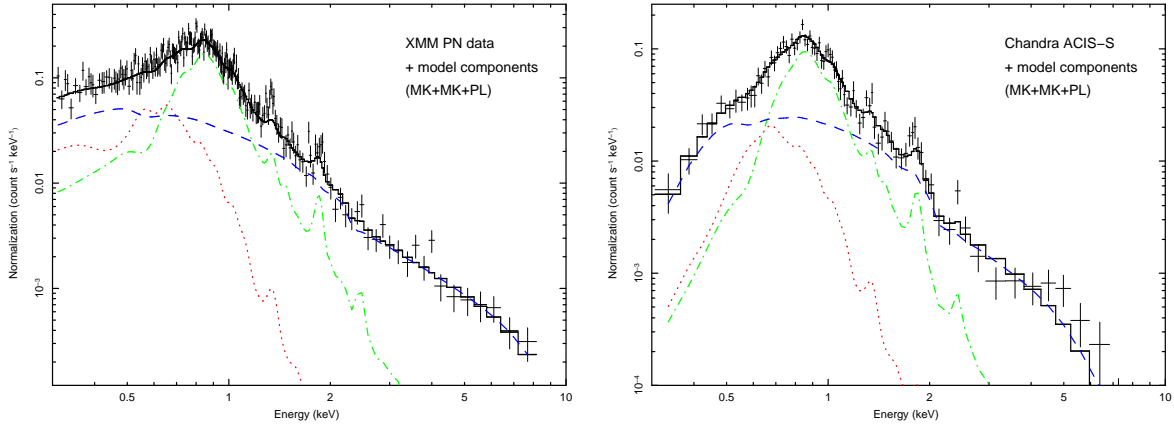

\begin{center}
\rotatebox{270}{\scalebox{0.32}{\includegraphics{fig3a.ps}}}
\rotatebox{270}{\scalebox{0.32}{\includegraphics{fig3b.ps}}}
\caption{X-ray spectra for the 20 arcsec radius central region of NGC~1672.  Left: {\it XMM-Newton} folded model and PN spectrum to illustrate the contributions from the model components, denoted by dashed (power-law), dotted (cool MEKAL) and dot-dashed (warm MEKAL) lines.  Right: the same for the ACIS-S (showing MEKAL model components frozen to the {\it XMM} fit values, but with a free power-law).}
\label{fig:spec_ctr}
\end{center}
\end{figure*}

\section{Nucleus}

As part of this program we have also obtained {\it HST}-ACS images of NGC~1672 in the F435W (B), F550M (V), F814W (I) and F658N (H$\alpha$) filters. Figure~\ref{fig:ctr} shows {\it Chandra} and {\it HST}/H$\alpha$ images of the central region of the galaxy.  For the first time, the {\it Chandra} data allow us to spatially resolve the X-ray emission into a weak point source surrounded by the circumnuclear ring.  The exquisite high-resolution {\it HST} image also demonstrates for the first time that NGC~1672 possesses a nuclear spiral.  The X-ray contours are overlaid on the H$\alpha$ image, and show a strong correlation between the location of the circumnuclear ring and the outer edge of the spiral.  The central X-ray source is also spatially co-incident with the peak of the H$\alpha$ emission. 

The central X-ray source is detected in the medium (1--2\,keV) and hard band (2--10\,keV) data ($\sim$30 counts in total), indicating that there is some obscuring material around the nucleus.  Hardness ratios indicate that this is a hard source, with a power-law photon index of $\Gamma\sim1.2$ and an absorbing column density up to N$_H\sim10^{23}$ cm$^{-2}$.  This translates into an unabsorbed 2--10\,keV luminosity of only $\sim1\times10^{39}$ erg s$^{-1}$; this is very faint for a Seyfert galaxy, but does just fall into the range found in the low-luminosity AGN survey of \cite{ho01} ($10^{38}-10^{41}$ erg s$^{-1}$).

Figure~\ref{fig:spec_ctr} ({\it left}) shows the {\it XMM-Newton} spectra of the 20 arcsec radius nuclear region, which includes all contributions from the nucleus and circumnuclear ring.  The best-fit model is comprised of two thermal MEKAL components ($kT$=0.2/0.6\,keV) and a soft power-law ($\Gamma\sim2.1$).  The 2--10\,keV luminosity of this model is $\sim2.7\times10^{39}$ erg s$^{-1}$.  Note that there is no evidence for a bright AGN-like component that would be absorbed at low energies and only transmitted at higher energies.  There is also no indication of a neutral Fe-6.4\,keV emission line, typically seen in bright Type-2 AGN (although such a line, if present, would likely be diluted by the emission from the starburst ring).  Also shown is the {\it Chandra} ACIS-S spectrum (Figure~\ref{fig:spec_ctr}, {\it right}), extracted from exactly the same nuclear region.  All model components are frozen to the {\it XMM} fit values except for the power-law.  This is slightly softer ($\Gamma\sim2.3$) than in the {\it XMM} observation, and has a 2--10\,keV luminosity of $\sim2.3\times10^{39}$ erg s$^{-1}$, consistent with the sum of the hard emission from the nucleus and point sources in the circumnuclear ring.  This is $\sim$20\% fainter than in the {\it XMM} data, but it is impossible to tell which of the sources in the central region may have varied over the $\sim$1.5 years between observations. 

The 0.2--2\,keV luminosity of the central source in the {\it ROSAT} HRI observations stayed constant at $\sim7\times10^{39}$ erg s$^{-1}$ between the 1992 and 1997 observations.  This agrees very well with both the {\it XMM} and {\it Chandra} 0.3--2\,keV emission (7.1 and 7.4$\times10^{39}$ erg s$^{-1}$ respectively);  there is therefore no sign of variability (in the soft band) over a $\sim$14 year period.

Other evidence of low-level AGN activity comes from radio imaging data, which shows that NGC~1672 has a compact 5\,GHz radio core located at its optical nucleus (see Jenkins et al. (2008) {\it in prep}). \cite{kewley00} also detect compact radio emission at 2.3\,GHz in the core of NGC~1672, but demonstrate that it is possible that this can be produced by clumps of luminous radio supernovae, similar to those found in the nuclei of Arp~220 and M82.  We also cannot rule out (at this time) that the hard X-ray emission may be produced by either one ULX or a few XRBs associated with star formation in the nuclear region, or that it may be a composite object with {\it some} of the emission coming from star-formation.  Other diagnostics such as the L$_X$/L$_{H\alpha}$ correlation of \cite{ho01} may provide further insight to the  nature of the central ionizing source.

\section{Summary \& Future Work}

With high-spatial-resolution {\it Chandra} imaging we have shown, for the first time, that NGC~1672 possesses a hard central X-ray source, with a low 2--10\,keV luminosity of $\sim10^{39}$ erg s$^{-1}$.   This in turn is surrounded by an X-ray bright circumnuclear star-forming ring, which dominates the X-ray emission in the central region of the galaxy.  

If the faint, hard X-ray source denotes low-level AGN activity, this could be an example of a much broader trend.  With similar studies on large samples of nearby galaxies, we will address the important question of the real nature of nuclear activity in barred versus non-barred systems.  For example, if detected, what range of X-ray luminosites and absorbing columns do these sources display?  What effect does a circumnuclear ring have on the luminosity, i.e. are they all low-luminosity because the ring may be stopping the bar-driven material from reaching the nucleus (as suggested by \citealt{ho97})?  This information will be a crucial element in assessing the overall demography of massive black holes in the nearby Universe.

\end{document}